\documentclass[a4paper,11pt]{article}
\usepackage{jcappub}

\usepackage{mathtools}

\newcommand{\rR}{\rho_R}
\newcommand{\rbh}{\rho_\text{BH}}
\newcommand{\gs}{g_\star}
\newcommand{\gss}{g_{\star s}}
\newcommand{\Teq}{T_\text{eq}}
\newcommand{\Tequal}{T_\text{equality}}
\newcommand{\Min}{M_\text{in}}
\newcommand{\Tin}{T_\text{in}}
\newcommand{\Tev}{T_\text{ev}}
\newcommand{\Tc}{T_\text{c}}
\newcommand{\Mbh}{M_\text{BH}}
\newcommand{\Tbh}{T_\text{BH}}
\newcommand{\Tosc}{T_\text{osc}}
\newcommand{\Tqcd}{T_\text{QCD}}

\title{ALP Dark Matter in a Primordial Black Hole Dominated Universe}

\author[a]{Nicolás Bernal,}
\author[b,c,d,e]{Yuber F.~Perez-Gonzalez,}
\author[f]{\\Yong Xu}
\author[g]{and Óscar Zapata}
\affiliation[a]{Centro de Investigaciones, Universidad Antonio Nariño\\Carrera 3 este \# 47A-15, Bogotá, Colombia}
\affiliation[b]{Theoretical Physics Department, Fermi National Accelerator Laboratory\\P.O. Box 500,
	Batavia, IL 60510, USA}
\affiliation[c]{Department of Physics \& Astronomy, Northwestern University, Evanston, IL 60208, USA}
\affiliation[d]{Colegio de Física Fundamental e Interdisciplinaria de las Américas (COFI)\\254 Norzagaray street, San Juan, Puerto Rico 00901}
\affiliation[e]{Institute for Particle Physics Phenomenology, Durham University\\South Road, Durham, U.K.}
\affiliation[f]{Bethe Center for Theoretical Physics and Physikalisches Institut, Universit\"at Bonn\\Nussallee 12, 53115 Bonn, Germany}
\affiliation[g]{Instituto de Física, Universidad de Antioquia\\Calle 70 \# 52-21, Apartado Aéreo 1226, Medellín, Colombia}

\emailAdd{nicolas.bernal@uan.edu.co}
\emailAdd{yuber.f.perez-gonzalez@durham.ac.uk}
\emailAdd{yongxu@th.physik.uni-bonn.de}
\emailAdd{oalberto.zapata@udea.edu.co}

\abstract{We investigate the phenomenological consequences of axion-like particle (ALP) dark matter with an early matter domination triggered by primordial black holes (PBHs). We focus on light BHs with masses smaller than $\sim 10^9~$g which fully evaporate before Big Bang nucleosynthesis. We numerically solve the coupled Boltzmann equations, carefully taking the greybody factors and BH angular momentum into account. We find that the entropy injection from PBH evaporation dilutes the ALP relic abundance originally produced via the vacuum misalignment mechanism, opening the parameter space with larger scales $f_a$ or, equivalently, smaller ALP-photon couplings $g_{a\gamma}$, within the reach of future detectors as ABRACADABRA, KLASH, ADMX, and DM-Radio. Moreover, the ALP minicluster masses can be several orders of magnitude larger if the early Universe features an PBH dominated epoch. For the relativistic ALPs produced directly from Hawking radiation, we find that their contribution to the dark radiation is within the sensitivity of next generation CMB experiments.
	For the sake of completeness, we also revisit the particular case of the QCD axion.
}

\begin{document}
	\begin{flushright}
		PI/UAN-2021-702FT, FERMILAB-PUB-21-478-T, NUHEP-TH/21-16,  IPPP/21/36
	\end{flushright}
	
	\maketitle
	
	\section{Introduction}
	Primordial black holes (PBHs), which could have been copiously produced in the early Universe due to large density fluctuations, are attracting intensive investigations, for recent reviews see e.g. Refs.~\cite{Carr:2020gox, Carr:2020xqk, Villanueva-Domingo:2021spv, Carr:2021bzv}.
	Those with masses larger than $\sim 10^{15}$~g have not fully evaporated at present and are potential cold dark matter (DM) candidates.
	On the contrary, light PBHs could have also been created, and have not a less interesting phenomenology.
	For example, in their evaporation process they can source particle DM or dark radiation (DR)~\cite{Hooper:2019gtx, Masina:2020xhk, Baldes:2020nuv, Gondolo:2020uqv, Bernal:2020kse, Bernal:2020ili, Bernal:2020bjf, Masina:2021zpu, Arbey:2021ysg, JyotiDas:2021shi, Cheek:2021odj, Cheek:2021cfe, Sandick:2021gew}, trigger baryogenesis~\cite{Barrow:1990he, Hamada:2016jnq, Hooper:2020otu, Perez-Gonzalez:2020vnz, Datta:2020bht}, and radiate all new degrees of freedom, such as right-handed neutrinos~\cite{Lunardini:2019zob}.
	Along the same research direction, in this work we focus on phenomenology of QCD axion and axion-like particles (ALPs) in a Universe dominated by PBHs lighter than $\sim 10^9$~g\footnote{For production mechanism of such light PBHs, see e.g. Ref.~\cite{Auclair:2020csm}.}, which have fully evaporated before Big Bang nucleosynthesis (BBN).
	
	The QCD axion~\cite{Weinberg:1977ma, Wilczek:1977pj}, a pseudo-Nambu-Goldstone boson resulting from the spontaneous breaking of the global $U(1)$ Peccei-Quinn symmetry~\cite{Peccei:1977hh, Peccei:1977np, Peccei:1977ur}, is a well-motivated candidate for DM. In the early Universe, it can be produced via a number of processes, being the vacuum misalignment mechanism~\cite{Preskill:1982cy, Abbott:1982af, Dine:1982ah} and the decay of topological defects~\cite{Sikivie:2006ni} the most popular in the literature.
	In the standard cosmological scenario, the axion window is rather narrow, in particular an axion with mass $m_a \simeq [10^{-6}-10^{-5}]$~eV (or correspondingly to a Peccei-Quinn scale $f_a \simeq 10^{12}$~GeV) is expected, if one does not want to introduce fine tuning of the initial misalignment angle $\theta_i$~\cite{Marsh:2015xka, DiLuzio:2020wdo, Sikivie:2020zpn}\footnote{We note that for very low scale inflation, axion field follows a  Bunch-Davies distribution, in which case $\theta_i$ could be samll \cite{Takahashi:2018tdu, Ho:2019ayl}.}. In light of the accumulated interesting implications for PBHs~\cite{Bird:2016dcv, Sasaki:2016jop, Clesse:2016vqa, Clesse:2017bsw, Gow:2019pok, Schiavone:2021imu}, it has been recently shown that the axion DM window can be enlarged to a mass as low as $m_a \sim \mathcal{O}(10^{-8})$~eV if the Universe features an early PBHs dominated epoch~\cite{Bernal:2021yyb}.
	
	In this paper, we go beyond the QCD axion DM and consider general ALPs, see e.g. Refs.~\cite{Jaeckel:2010ni, Arias:2012az, Ringwald:2012hr, Marsh:2015xka, Irastorza:2018dyq, DiLuzio:2020wdo}. 
	ALPs could also arise from the spontaneous breaking of a global $U(1)$ symmetry, similar to the QCD axion. They are  also quite ubiquitous in string theory~\cite{Svrcek:2006yi, Arvanitaki:2009fg}. Differently from QCD axions, ALPs do not solve the strong CP problem since they are in general not involved in the strong interaction. However, they serve as good candidate for DM, see e.g. Ref.~\cite{Ringwald:2012hr}. In this paper, we focus on ALP DM generated via the usual misalignment mechanism. Therefore, our results are complementary to those presented in Ref.~\cite{Schiavone:2021imu}.
	In the standard cosmological scenario, the ALP relic density depends on three parameters: the decay constant $f_a$, its mass $m_a$ and the initial misalignment angle $\theta_i$, leading to strong bounds on the viable parameter space for natural values of $\theta_i$.
	However, new regions of the parameter space become viable with a nonstandard cosmological epoch before BBN.
	We note that the phenomenology of ALPs as DM with nonstandard cosmological epochs such as kination or an early matter phase was investigated in Ref.~\cite{Blinov:2019rhb} recently, where it was shown that a broader parameter is available and could be within the reach of several proposed experiments. 
	Similarly to Ref.~\cite{Bernal:2021yyb}, and differently from Ref.~\cite{Blinov:2019rhb}, we focus on the phenomenological consequences of ALPs in an early matter epoch triggered by PBHs.
	
	Since PBHs effectively behave as matter, with an energy density that redshifts slower than radiation, it could be expected that the early Universe has undergone a PBH dominated epoch.
	Moreover, it is worth to mention some differences between an early matter dominated epoch triggered by PBHs and by a long-lived heavy particle. In the former case, the evaporation (or decay) rate is time dependent while the later is not. Besides, an early PBH era would inevitably give rise to gravitational wave (GW) signatures~\cite{Inomata:2020lmk,Papanikolaou:2020qtd, Domenech:2020ssp, Domenech:2021wkk, Domenech:2021ztg}, offering an interesting avenue to constrain on the initial amount of energy density in the early matter epoch, while such kind of constraints are absent in case of a heavy particle.
	
	For the case with an early PBH era, we numerically solve a system of coupled Boltzmann equations for the background (PBHs and SM radiation) based on the code developed in Refs.~\cite{Cheek:2021odj, Cheek:2021cfe}, carefully including the greybody factor in PBHs spectra, and considering the effect of the PBH angular momentum.  We find an increased allowed parameter space with larger $f_a$ or, equivalently, smaller ALP-photon coupling $g_{a\gamma}$ due to the entropy injection from PBH evaporation. 
	Furthermore, in the scenario where the Peccei-Quinn symmetry is spontaneously broken after inflation, the gravitational clump of ALP (axion) density inhomogeneities at the time of the matter-radiation equality gives rise to miniclusters~\cite{Hogan:1988mp, Kolb:1993hw, Kolb:1993zz, Kolb:1994fi, Eggemeier:2019khm}.  Due to the PBH domination epoch, miniclusters with larger masses could be formed.
	
	Additionally, ALPs and axions are directly radiated from PBHs evaporation, being ultra-relativistic, and thus contributing to the DR.
	Taking carefully into account the effect of the PBH spin, we numerically compute the contribution to $\Delta N_{\text{eff}}$, which is within the sensitivity of next generation CMB experiments~\cite{CMB-S4:2016ple, NASAPICO:2019thw, Benson:2014qhw, Ade:2018sbj}. We also find that for Kerr  PBHs,  $\Delta N_{\text{eff}}$ turns out to be smaller than for nonrotating Schwarzschild PBHs.
	
	The paper is organized as follows. We first revisit PBH evaporation for both Schwarz\-schild and Kerr BHs in Sec.~\ref{PBH_eva}. In Sec.~\ref{Beq_entropy}, we estimate the background evolution by setting up the coupled Boltzmann equations and the formalism for estimating the entropy injection. Sec.~\ref{ALP} is devoted to the phenomenological consequences of ALP DM. For the sake of completeness, in Sec.~\ref{axion_DM} we revisit QCD axion DM. Additionally, in Sec.~\ref{axion_DR} we focus on the axion and ALP DR directly produced from PBH evaporation. Finally we sum up our findings in Sec.~\ref{conclusion}. We use natural units where $\hbar = c = k_{\rm B} = 1$ throughout this manuscript. 
	
	\section{Primordial Black Hole Evaporation} \label{PBH_eva}
	PBHs are hypothetical objects which could be generated due to inhomogeneities of density perturbations in the early Universe~\cite{Hawking:1975vcx, Carr:1974nx}. When these fluctuations reenter the horizon, if above a threshold, they could collapse and form a BH according to the Press-Schechter formalism~\cite{Press:1973iz}. In this paper, we focus on the case where PBHs form in a radiation-dominated epoch.  With an initial cosmic temperature $T = \Tin$, the initial PBH mass is given by the whole mass within the particle horizon~\cite{Carr:2009jm, Carr:2020gox},
	\begin{equation}\label{eq:Mi}
	\Min \equiv \Mbh(\Tin) = \frac{4\pi}{3}\, \gamma\, \frac{\rR(\Tin)}{H^3(\Tin)}\,,
	\end{equation}
	where we take the efficiency factor to be $\gamma \simeq 0.2$.
	Additionally, $\rR$ and $H$ correspond to the standard model (SM) energy density and the Hubble expansion rate, respectively.
	For simplicity, we assume a monochromatic mass spectrum, such that all PBHs were created with the same mass. 
	The extension to more realistic mass distributions will be considered elsewhere.
	The PBH initial energy density $\rbh(\Tin)$ is usually related to the SM radiation energy density at formation via the $\beta$ parameter
	\begin{equation}
	\beta \equiv \frac{\rbh(\Tin)}{\rR(\Tin) }\,.
	\end{equation}
	Since $\rbh$ redshifts slower than radiation, an early PBH-dominated epoch with $\rbh > \rR$ is triggered if $\beta > \beta_c$, with
	\begin{equation}
	\beta_c \equiv \frac{\Tev}{\Tin}\,,
	\end{equation}
	where $\Tev$, given by~\cite{Bernal:2021yyb}
	\begin{equation}
	\Tev \simeq \left(\frac{\gs(\Tin)}{640}\right)^{1/4} \left(\frac{M_P^5}{\Min^3}\right)^{1/2},
	\end{equation}
	is the SM temperature at which PBHs completely evaporate.
	Additionally, $M_P  \equiv 1/\sqrt{8\pi\, G}$ denotes the reduced Planck mass, and $\gs(T)$ corresponds to the number of relativistic degrees of freedom contributing to the SM energy density.
	It is interesting to note that if $\beta > \beta_c$, PBHs start to dominate the total energy density of the Universe at $T = \Teq$, defined as $\rR(\Teq) = \rbh(\Teq)$, with
	\begin{equation}
	\Teq = \beta\, \Tin \left(\frac{\gss(\Tin)}{\gss(\Teq)}\right)^{1/3},
	\end{equation}
	with $\gss(T)$ being the number of relativistic degrees of freedom contributing to the SM entropy.
	
	Several bounds exist in the PBH parameter space spanned by the initial fraction $\beta$ and mass $\Min$~\cite{Carr:2020gox, Carr:2020xqk}. 
	Here, we will focus on PBHs that evaporated before BBN, thus having masses smaller than $\sim 10^9$~g. 
	Although there exist constraints on such light PBHs, they are typically model dependent~\cite{Carr:2020gox, Carr:2020xqk, Lunardini:2019zob}. 
	Nevertheless, recent constraints have been derived after considering the GWs  emitted from the Hawking evaporation.
	In particular, a backreaction problem can be avoided if the energy contained in GWs never overtakes the one of the background universe~\cite{Papanikolaou:2020qtd}.
	More importantly, a modification of BBN predictions due to the energy density stored in GWs can be avoided if~\cite{Domenech:2020ssp}
	\begin{equation} \label{eq:GW}
	\beta \lesssim 1.1 \times 10^{-6} \left(\frac{\gamma}{0.2}\right)^{-\frac12} \left(\frac{\Min}{10^4~\text{g}}\right)^{-\frac{17}{24}}.
	\end{equation}
	
	Next, we briefly describe the time evolution properties of Schwarzschild and Kerr PBHs relevant for our purposes. 
	Further details can be found in Refs.~\cite{Page:1976df, Cheek:2021cfe, Cheek:2021odj}.
	
	\subsection{Schwarzschild Black Holes}
	For the most simple BH scenario, PBHs are only characterized by their mass, so that their horizon temperature $\Tbh$ is determined according to~\cite{Hawking:1975vcx}
	\begin{equation}
	\Tbh = \frac{M_P^2}{\Mbh} \simeq 10^{13}\,{\rm GeV} \left(\frac{1\,{\rm g}}{\Mbh}\right).
	\end{equation}
	All particles with masses smaller than $\Tbh$ can be emitted via Hawking radiation.  Within a time interval $dt$ and momentum $[p,\, p + dp]$, the energy spectrum of an emitted species $i$ with spin $s_i$, mass $\mu_i$ and internal degrees of freedom $g_i$ can be described as~\cite{Hawking:1975vcx, Cheek:2021odj}
	\begin{equation} \label{eq:spectrum}
	\frac{\text{d}^{2} \mathcal{N}_{i}}{\text{~d}p~\text{d}t} = \frac{g_{i}}{2 \pi^{2}}\, \frac{\sigma_{s_{i}}\left(\Mbh, \mu_{i}, p\right)}{\exp \left[E_{i}(p) / T_{\text{BH}}\right]-(-1)^{2 s_{i}}}\, \frac{p^{3}}{E_{i}(p)}\,,
	\end{equation}
	with $E_i(p) = \sqrt{p^2 + \mu_i^2}$ denoting the energy and $\sigma_{s_i}$ describing the absorption cross-section. Summing over all possible particle species and integrating over the phase space in Eq.~\eqref{eq:spectrum}, one can obtain the BH mass evolution, which is shown to be~\cite{MacGibbon:1990zk, MacGibbon:1991tj}
	\begin{equation} \label{eq:Schwarzschild}
	\frac{\text{d} \Mbh}{\text{d} t} \equiv \sum_{i} \left.\frac{\text{d} \Mbh}{\text{d} t}\right|_{i}=-\sum_{i} \int_{0}^{\infty} E_{i} \frac{\text{d}^{2} \mathcal{N}_{i}}{\text{~d} p \text{~d} t} \text{~d} p =-\varepsilon\left(\Mbh\right) \frac{M_P^{4}}{\Mbh^{2}}\,,
	\end{equation}
	with the mass evaporation function $\varepsilon\left(\Mbh\right) \equiv \sum_i g_{i}\, \varepsilon_{i}(z_i)$, where the contribution per degree of freedom $\varepsilon_{i}(z_i)$ is given by~\cite{Cheek:2021odj}
	\begin{equation}
	\varepsilon_{i}(z_i) = \frac{27}{128 \pi^{3}} \int_{z_{i}}^{\infty} \frac{\psi_{s_{i}}\left(x, z_{i}\right)\left(x^{2}-z_{i}^{2}\right)}{\exp (x)-(-1)^{2 s_{i}}}\, x \text{~d} x\,,
	\end{equation}
	where $x = E_i / \Tbh$, $z_i = \mu_i / \Tbh$, and $\psi_{s_{i}}(E,\mu) \equiv \sigma_{s_{i}}(E,\mu)/(27\pi\, G^2\, \Mbh^2)$.
	
	\subsection{Kerr Black Holes}
	For the case of BHs with a nonzero spin, the horizon temperature has an additional dependence on the spin parameter $a_\star \in [0,1]$ (with $a_\star = 0$ corresponding to the Schwarzschild limit)
	\begin{equation}
	\Tbh = \frac{2M_P^2}{\Mbh}\, \frac{\sqrt{1-a_\star}}{1+\sqrt{1-a_\star}}\,,
	\end{equation}
	where $a_\star \equiv 8\pi\, J\, M_P^2/\Mbh^2$, with $J$ denoting the total BH angular momentum. The energy spectrum is modified by the presence of the angular momentum, introducing an explicit dependence on the total $l$ and axial $m$ angular quantum numbers,
	\begin{equation}
	\frac{\text{d}^{2} \mathcal{N}_{i}}{\text{~d} p \text{~d} t}=\frac{g_{i}}{2 \pi^{2}} \sum_{l=s_{i}} \sum_{m=-l}^{l} \frac{\sigma_{s_{i}}^{l m}\left(\Mbh, p, a_{\star}\right)}{\exp \left[\left(E_{i}-m\, \Omega\right) / T_{\text{BH}}\right]-(-1)^{2 s_{i}}}\, \frac{p^{3}}{E_{i}}\,,
	\end{equation}
	with $\Omega = \left(4\pi\, a_{\star}\, M_P^2/\Mbh\right) \left(1+\sqrt{1-a_{\star}^{2}}\right)^{-1}$ horizon angular velocity. 
	The time evolution for the BH spin and mass is described by the following set of equations~\cite{Page:1976df} 
	\begin{subequations} \label{eq:eveqsKerr}
		\begin{align}
		\frac{d\Mbh}{dt} &= -\varepsilon\left(\Mbh, a_{\star}\right) \frac{M_P^{4}}{\Mbh^{2}}\,, \\
		\frac{d a_{\star}}{d t} &=-a_{\star}\left[\gamma\left(\Mbh, a_{\star}\right)-2 \varepsilon\left(\Mbh, a_{\star}\right)\right] \frac{M_P^{4}}{\Mbh^{3}}\,,
		\end{align}
	\end{subequations}
	where the evaporation functions $\varepsilon\left(\Mbh, a_{\star}\right)$ and $\gamma\left(\Mbh\right)$ are given by
	\begin{subequations} \label{eq:epgK}
		\begin{align}
		\varepsilon_{i}\left(z_{i}, a_{\star}\right)&=\frac{27}{128 \pi^{3}} \int_{z_{i}}^{\infty} \sum_{l, m} \frac{\psi_{s_{i}}^{l m}\left(x, a_{\star}\right)\left(x^{2}-z_{i}^{2}\right)}{\exp \left(x^{\prime} / 2 f\left(a_{\star}\right)\right)-(-1)^{2 s_{i}}}\,  x \text{~d} x\,,\\
		\gamma_{i}\left(z_{i}, a_{\star}\right) &= \frac{27}{16 \pi^{2}} \int_{z_{i}}^{\infty} \sum_{l, m} \frac{m\, \psi_{s_{i}}^{l m}\left(x, a_{\star}\right)\left(x^{2}-z_{i}^{2}\right)}{\exp \left(x^{\prime} / 2 f\left(a_{\star}\right)\right)-(-1)^{2 s_{i}}}\,  \text{d} x\,,
		\end{align}
	\end{subequations}
	with $x^{\prime} = x-m\Omega^\prime$, being\footnote{Note that the definition of $x$ is the same as in the Schwarzschild case.} $x=\Mbh E_i/M_P^2$, $\Omega^\prime = \Mbh \Omega/M_P^2$, and $f(a_\star) = \sqrt{1 - a_\star}/(1 + \sqrt{1 - a_\star})$.
	This set of time evolution equations are the basis for describing the effect of a PBH dominated Universe on the ALP and axion DM genesis.
	
	\section{Background Evolution} \label{Beq_entropy}
	The evolution of the SM entropy density  $s(T) = \frac{2\pi^2}{45} \gss(T)\, T^3$ can be tracked via the Boltzmann equation
	\begin{equation} 
	\frac{ds}{dt} + 3\, H\,s = -\frac{1}{T}\, \frac{\rbh}{\Mbh} \left.\frac{d\Mbh}{dt}\right|_\text{SM}, \label{eq:BE2}
	\end{equation}
	where $H^2 = (\rR + \rbh)/(3M_P^2)$, with $\rR(T) = \frac{\pi^2}{30} \gs(T)\, T^4$, and $\frac{-1}{\Mbh} \left.\frac{d\Mbh}{dt}\right|_\text{SM}$ corresponds to the {\it time-dependent} PBH evaporation rate into SM particles.
	We emphasize that Eq.~\eqref{eq:BE2} has to be numerically solved together with Eq.~\eqref{eq:Schwarzschild} in the case of a Schwarzschild PBH, or with Eqs.~\eqref{eq:eveqsKerr} in the case of a Kerr PBH, to extract the dynamics of the background, and in particular the evolution of the SM temperature and the (non-conserved) SM entropy.
	
	In a SM radiation dominated scenario, the Hubble expansion rate takes the simple form
	\begin{equation}
	H(T) = H_R(T) \equiv \sqrt{\frac{\rR(T)}{3\, M_P^2}} = \frac{\pi}{3} \sqrt{\frac{\gs(T)}{10}}\, \frac{T^2}{M_P}\,.
	\end{equation}
	However, PBHs can have a strong impact on the evolution of the background dynamics~\cite{Allahverdi:2020bys}.
	With an early PBH dominated epoch, cosmology can be characterized by four distinct regimes, where the Hubble expansion rate is given by~\cite{Bernal:2021yyb}
	\begin{equation} \label{eq:H1}
	H(T) \simeq
	\begin{dcases}
	H_R(T) & \text{for } T \geq \Teq\,,\\
	H_R(\Teq) \left[\frac{\gss(T)}{\gss(\Teq)} \left(\frac{T}{\Teq}\right)^3\right]^{1/2} & \text{for } \Teq \geq T \geq \Tc\,,\\
	H_R(\Tev) \left[1 - \frac{720}{\pi\, \gs(\Tin)} \frac{\Min^3}{M_P^4} \frac{H_R^2(\Tev) - H_R^2(T)}{H_R(\Tev)}\right] & \text{for }  \Tc \geq T \geq \Tev\,,\\
	H_R(T) & \text{for } \Tev \geq T\,,
	\end{dcases}
	\end{equation}
	which can be simplified to
	\begin{equation}
	H(T) \simeq
	\begin{dcases}
	H_R(T) & \text{ for } T \geq \Teq\,,\\
	H_R(\Teq) \left(\frac{T}{\Teq}\right)^{3/2} & \text{ for } \Teq \gg T \gg \Tc\,,\\
	H_R(\Tev) \left(\frac{T}{\Tev}\right)^4 & \text{ for } \Tc \gg T \gg \Tev\,,\\
	H_R(T) & \text{ for } \Tev \geq T\,,
	\end{dcases}
	\end{equation}
	for being more conveniently used in next expressions.
	Here, we have introduced the temperature scale $\Tc$, from which the SM radiation does not scale as free radiation (i.e., $\rR(R) \propto R^{-1}$ with $R$ being the scale factor) due to the entropy injected by the PBH evaporation.
	It is given by~\cite{Bernal:2021yyb}
	\begin{equation}
	\Tc \simeq \left[\frac{\gs(\Tin)\, \pi}{5760}\, \frac{M_P^{10}\, \Teq}{\Min^6}\right]^{1/5}
	\simeq \left(\Teq\, \Tev^4\right)^{1/5}.
	\end{equation}
	
	Additionally, during its evaporation, PBHs radiate SM particles and therefore dilute all previously produced species.
	The entropy injection factor is~\cite{Bernal:2021yyb}
	\begin{equation} \label{eq:S1}
	\frac{S(T)}{S(\Tev)} \simeq
	\begin{dcases}
	\frac{\gss(\Teq)}{\gss(\Tev)}\, \frac{\gs(\Tev)}{\gs(\Teq)}\, \frac{\Tev}{\Teq} & \text{for } T \geq \Tc\,,\\
	\frac{\gss(T)}{\gss(\Tev)} \left(\frac{T}{\Tev}\right)^3 \left[1 - \frac{720}{\pi\, \gs(\Tin)} \frac{\Min^3}{M_P^4} \frac{H_R^2(\Tev) - H_R^2(T)}{H_R(\Tev)}\right]^{-2} & \text{for } \Tc \geq T \geq \Tev\,,\\
	1 & \text{for } \Tev \geq T\,,
	\end{dcases}
	\end{equation}
	which can be simplified to
	\begin{equation} \label{eq:S2}
	\frac{S(T)}{S(\Tev)} \simeq
	\begin{dcases}
	\frac{\Tev}{\Teq} & \text{ for } T \geq \Tc\,,\\
	\left(\frac{\Tev}{T}\right)^5 & \text{ for } \Tc \geq T \geq \Tev\,,\\
	1 & \text{ for } \Tev \geq T\,,
	\end{dcases}
	\end{equation}
	again to ease further analytical estimations.
	We note that in Eqs.~\eqref{eq:S1} and~\eqref{eq:S2} there are three distinctive regimes, and not four as in Eq.~\eqref{eq:H1}.
	
	Finally, it is important to emphasize that all these analytical estimations help to understand the dynamics of the cosmological evolution.
	However, hereafter full numerical solutions of the background will be used.
	In our code, we solve such Boltzmann equations together with the time evolution equations for the PBHs, given in Eqs.~\eqref{eq:eveqsKerr}. 
	Notice that taking as initial value $a_\star=0$, the evolution equations in~\eqref{eq:eveqsKerr} reduce to the ones for the Schwarzschild scenario, i.e., Eq.~\eqref{eq:Schwarzschild}. 
	In other words, such scenario is included in the more general Kerr case. 
	Thus, we take in general the equations for the Kerr case, putting  $a_\star=0$ whenever we talk about Schwarzschild PBHs.
	Besides, in the evaporation functions $\varepsilon\left(\Mbh, a_{\star}\right)$ and $\gamma\left(\Mbh\right)$ we have included an additional, almost massless, pseudoscalar degree of freedom, corresponding to the ALP or the axion.
	
	\section{ALP Dark Matter} \label{ALP}
	In this section, we focus on a general light pseudoscalar, namely the ALP~\cite{Jaeckel:2010ni, Arias:2012az, Ringwald:2012hr}.
	Similarly to the QCD axion, ALPs could arise as consequence of the spontaneous breaking of a global $U(1)$ symmetry or, alternatively, they could emerge from string theory~\cite{Svrcek:2006yi, Arvanitaki:2009fg}.
	
	In the so-called misalignment mechanism, the present energy density stored in the zero mode of an ALP field $a$ of mass $m_a$ can be obtained by solving the equation of motion~\cite{Preskill:1982cy, Stecker:1982ws, Abbott:1982af, Dine:1982ah}
	\begin{equation} \label{eom}
	\ddot \theta + 3\, H(t)\, \dot \theta + m_a^2\, \sin\theta = 0\,,
	\end{equation}
	where $\theta(t) \equiv a(t)/f_a$, and $f_a$ is the Peccei-Quinn (PQ) symmetry breaking energy scale.
	In this scenario, the  ALP  field  starts  to  roll  about  the  minimum  of  the potential once the Hubble friction is overcome by the potential term.
	Coherent oscillations of the ALPs are set around the temperature $\Tosc$ defined as
	\begin{equation}
	3\,H(\Tosc) \equiv m_a\,.
	\end{equation}
	
	In the standard cosmological scenario, as the SM entropy is conserved, the ALP energy density $\rho_a$ at present is given by
	\begin{equation} \label{rhoALP1}
	\rho_a(T_0) = \rho_a(\Tosc)\, \frac{s(T_0)}{s(\Tosc)}\,,
	\end{equation}
	where $T_0$ is the CMB temperature at present, and we considered the conservation of the ALP number density in a comoving volume.
	Additionally, $\rho_a(\Tosc) \simeq \frac12  m_a^2\, f_a^2\, \theta_i^2$  in the limit in which the kinetic energy is neglected, and for a quadratic potential. Moreover, $\theta_i$ denotes the initial misalignment angle.
	The entropy injection from PBHs evaporation would inevitably dilute the energy density  so that one has
	\begin{equation} \label{eq:rhoALP2}
	\rho_a(T_0) = \rho_a(\Tosc)  \frac{s(T_0)}{s(\Tosc)} \times \frac{S(\Tosc)}{S(T_\text{ev})} \simeq \frac12 m_a^2\, f_a^2\, \theta_i^2\, \frac{s(T_0)}{s(\Tosc)} \times \frac{S(\Tosc)}{S(T_\text{ev})}\,,
	\end{equation} 
	where $S(T) = s(T)\, R^3$ corresponds to the total SM entropy. With such an energy density, one can compute the ALP DM relic density via 
	\begin{equation}
	\Omega_ah^2  \equiv \frac{\rho_a(T_0)}{\rho_c/h^2}\,,
	\end{equation} 
	where $\rho_c/h^2 \simeq 1.1 \times 10^{-5}$~GeV/cm$^3$ is the critical energy density and $s(T_0) \simeq 2.69 \times 10^3$~cm$^{-3}$ is the entropy density at present, which should match the total DM abundance $\Omega h^2 \simeq 0.12$~\cite{Planck:2018vyg}.
	
\begin{figure}
		\def\sepf{0.51}
		\centering
		\includegraphics[scale=\sepf]{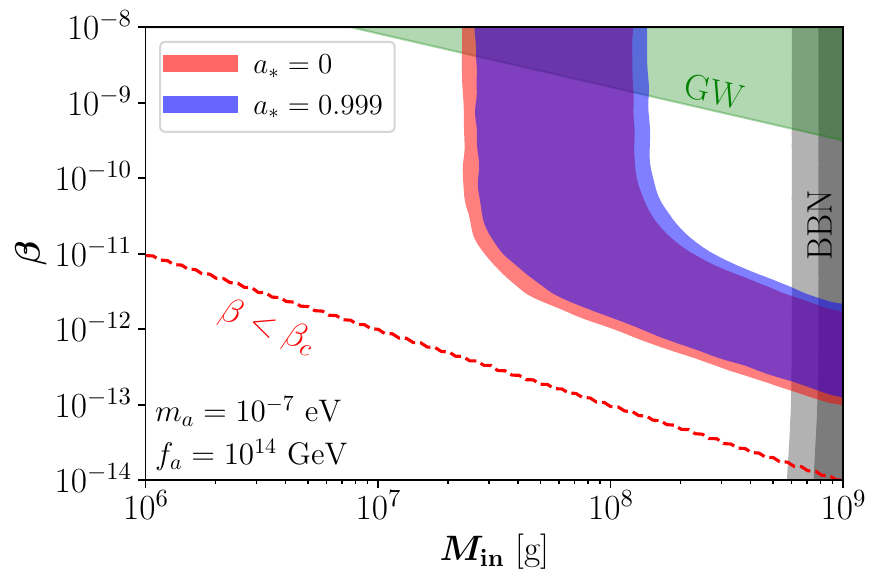}
		\caption{Parameter space compatible with the whole observed ALP DM abundance, for $m_a = 10^{-7}$~eV and $f_a = 10^{14}$~GeV.
			The thickness of the bands correspond to $0.5 \leq \theta_i \leq \pi/\sqrt{3}$. The red bands show $a_\star=0$, whereas the blue $a_\star=0.999$.
		}
		\label{fig:ALP}
	\end{figure}

	Knowing the Hubble expansion rate given in the previous section, the ALP oscillation temperature $\Tosc$ can be estimated to be
	\begin{equation} \label{eq:tosc}
	\Tosc \simeq
	\begin{dcases}
	\left(\frac{1}{\pi} \sqrt{\frac{10}{\gs}}\, M_P\, m_a\right)^{1/2} & \text{for } T \geq \Teq\,,\\
	\left(\frac{1}{\pi} \sqrt{\frac{10}{\gs}}\, \frac{M_P\, m_a}{\sqrt{\Teq}}\right)^{2/3} & \text{for } \Teq \geq T \geq \Tc\,,\\
	\left(\frac{1}{\pi} \sqrt{\frac{10}{\gs}}\, M_P\, m_a\, \Tev^2\right)^{1/4} & \text{for }  \Tc \geq T \geq \Tev\,,\\
	\left(\frac{1}{\pi} \sqrt{\frac{10}{\gs}}\, M_P\, m_a\right)^{1/2} & \text{for } \Tev \geq T\,.
	\end{dcases}
	\end{equation}
	Using Eq.~\eqref{eq:rhoALP2}, one can compute the relic density in the aforementioned four regimes, which becomes
	\begin{equation} \label{eq:Omegah2ALP}
	\Omega_a h^2 \simeq  \frac{45\, \theta_i^2}{4\pi^2\, \gss}\, \frac{s(T_0)}{\rho_c/h^2} \times
	\begin{dcases}
	\left(\sqrt{\frac{\gs}{10}}\, \frac{\pi}{M_P}\right)^{3/2} \frac{\Tev}{\Teq}\, m_a^{1/2}\, f_a^2 & \text{ for } \Tosc \geq \Teq\,,\\
	\left(\sqrt{\frac{\gs}{10}}\, \frac{\pi}{M_P}\right)^2 \Tev\, f_a^2 & \text{ for } \Teq \geq \Tosc \geq \Tev\,,\\
	\left(\sqrt{\frac{\gs}{10}}\, \frac{\pi}{M_P}\right)^{3/2} m_a^{1/2}\, f_a^2 & \text{ for } \Tev \geq \Tosc\,,
	\end{dcases}
	\end{equation}
	or equivalently
	{\small
		\begin{equation}
		\frac{\Omega_a h^2}{0.12} \simeq
		\begin{dcases}
		\left(\frac{\theta_i}{0.5}\right)^2 \left(\frac{\Tev}{4~\text{MeV}}\right) \left(\frac{81~\text{GeV}}{\Teq}\right) \left(\frac{m_a}{10^{-3}~\text{eV}}\right)^{1/2} \left(\frac{f_a}{5 \times 10^{14}~\text{GeV}}\right)^2 & \text{for } \Tosc \geq \Teq,\\
		\left(\frac{\theta_i}{0.5}\right)^2 \left(\frac{\Tev}{4~\text{MeV}}\right)\, \left(\frac{f_a}{10^{15}~\text{GeV}}\right)^2 & \text{for } \Teq \geq \Tosc \geq \Tev,\\
		\left(\frac{\theta_i}{0.5}\right)^2 \left(\frac{m_a}{10^{-15}~\text{eV}}\right)^{1/2}\, \left(\frac{f_a}{3\times 10^{15}~\text{GeV}}\right)^2 & \text{for } \Tev \geq \Tosc.
		\end{dcases}
		\end{equation}
	}
	Note that the expressions for $\Omega_a h^2$ are identical for  $\Teq \geq \Tosc \geq \Tc$ and $\Tc \geq \Tev$.
	Considering the full numerical solution of the Boltzmann equations, we present in Fig.~\ref{fig:ALP} the parameter space generating the whole observed ALP DM abundance for benchmark values of $m_a = 10^{-7}$~eV and $f_a = 10^{14}$~GeV, while assuming Schwarzschild ($a_\star=0$, red band) or Kerr ($a_\star=0.999$, blue band) PBHs.
	The thickness of the bands corresponds to an initial misalignment angle $0.5 \leq \theta_i \leq \pi/\sqrt{3}$.%
	\footnote{The lower bound comes from assuming an $\mathcal{O}(1)$ misalignment angle that avoids introducing fine tuning of the initial conditions, whereas the upper bound is an effective average misalignment angle, typical from post-inflationary scenarios.}
	The regions on the right (left) of the bands produce a DM underabundance (overabundance).
	Additionally, the gray regions are in tension with BBN ($\Tev < 4$~MeV~\cite{Sarkar:1995dd, Kawasaki:1999na, Kawasaki:2000en, Hannestad:2004px, deSalas:2015glj, Hasegawa:2019jsa}), whereas the green region with GWs (i.e., Eq.~\eqref{eq:GW}).
	Finally, the dashed red line corresponds to $\beta = \beta_c$, limiting the region where PBH energy density is subdominant with respect to SM radiation, and therefore one has a standard cosmological scenario (below), with the region where a nonstandard cosmological expansion triggered by PBHs occurs (above).
	It is important to emphasize that these results where obtained using the numerical code developed in Ref.~\cite{Cheek:2021cfe, Cheek:2021odj}, including the additional ALP degree of freedom.
	We have checked that the numerical estimations in Eq.~\eqref{eq:Omegah2ALP} fit well the analytical results.
	
\begin{figure}
		\def\sepf{0.51}
		\centering
		\includegraphics[scale=\sepf]{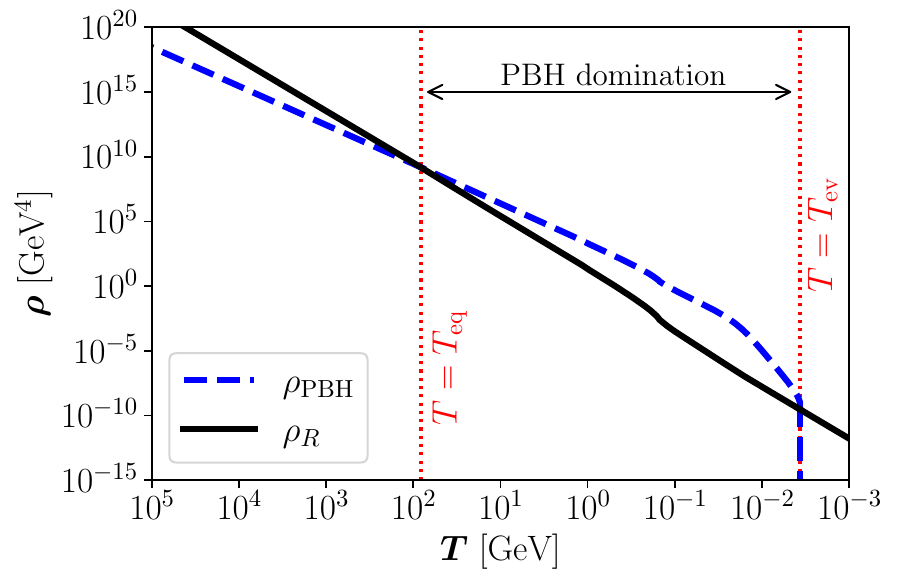}
		\includegraphics[scale=\sepf]{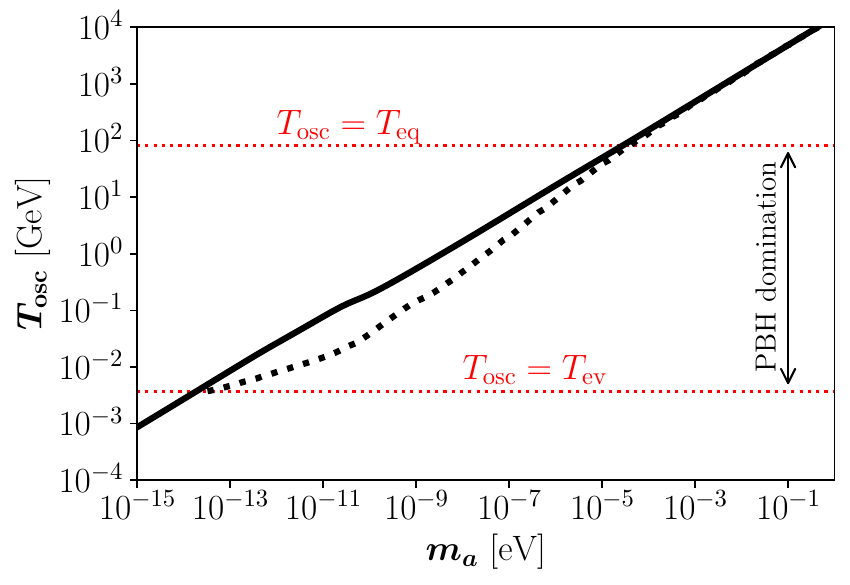}
		\caption{Left: Evolution of the energy densities for SM radiation and PBHs.
			Right: Oscillation temperature of ALPs, for radiation domination (solid) and PBH domination (dotted).
			For both panels, the maximal allowed deviation from the standard cosmological scenario for Schwarzschild PBHs was assumed, i.e. $\Min \simeq 5.7 \times 10^8$~g and $\beta \simeq 4.6 \times 10^{-10}$.}
		\label{fig:rho}
	\end{figure}

	As seen in Fig.~\ref{fig:ALP}, the presence of an era dominated by PBHs can have a strong impact on the ALP production in the early Universe.
	The maximal deviation from the standard cosmological scenario corresponds to long-lived PBHs evaporating just before the onset of BBN, and the maximal value for $\beta$ allowed by GWs.
	This happens for $\Min \simeq 5.7 \times 10^8$~g and $\beta \simeq 4.6 \times 10^{-10}$ for $a_\star = 0$, or $\Min \simeq 7.6 \times 10^8$~g and $\beta \simeq 3.5 \times 10^{-10}$ for $a_\star = 0.999$, and corresponds to the upper right white corners in Fig.~\ref{fig:ALP}.
	This maximal deviation from the standard cosmological scenario is shown in Fig.~\ref{fig:rho}, for the benchmark $\Min \simeq 5.7 \times 10^8$~g and $\beta \simeq 4.6 \times 10^{-10}$, for $a_\star = 0$.
	The left panel shows the evolution of the radiation and PBH energy densities as a function of the SM temperature.
	PBHs dominate the total energy density between $T = \Teq \simeq 81$~GeV and $T = \Tev \simeq 4$~MeV.
	Additionally, the right panel shows the oscillation temperature for radiation domination (solid), and PBH domination (dotted).
	As expected from Eq.~\eqref{eq:tosc}, for a given ALP mass, $\Tosc$ decreases for a PBH dominated era.
	Again, these results were obtained using the numerical code and fit well the analytical estimations.
	
	It is customary to assume that the global PQ symmetry is also anomalous respect to the electromagnetic gauge group thank to the existence of exotic vector-like charged fermions.
	Consequently, ALPs will couple to two photons, via the effective dimension-5 operator
	\begin{equation}
	\mathcal L_{a\gamma} = -\frac14\, g_{a\gamma}\, a\, F_{\mu\nu}\, \tilde F^{\mu\nu} = g_{a\gamma}\, a\, \vec E \cdot \vec B\,,
	\end{equation}
	where the coupling constant $g_{a\gamma}$ is model dependent and related to the breaking scale of the PQ symmetry as
	\begin{equation}
	g_{a\gamma} = \frac{\alpha}{2\pi\, f_a} \left(\frac{E}{N} - \frac23\, \frac{4+z}{1+z} \right) \simeq 10^{-13}~\text{GeV}^{-1} \left(\frac{10^{10}~\text{GeV}}{f_a}\right), 
	\end{equation}
	with $z \equiv m_u/m_d$, and $E$ and $N$ are the electromagnetic and color anomalies associated with the ALP anomaly. For KSVZ models $E / N = 0$~\cite{Kim:1979if, Shifman:1979if}, whereas for DFSZ models $E / N = 8 / 3$~\cite{Zhitnitsky:1980tq, Dine:1981rt}. 
	The electromagnetic interaction of ALPs is by far the most exploited to look for signatures in observations and experimental searches~\cite{Graham:2015ouw}.
	
\begin{figure}
		\def\sepf{0.40}
		\centering
		\includegraphics[scale=0.40]{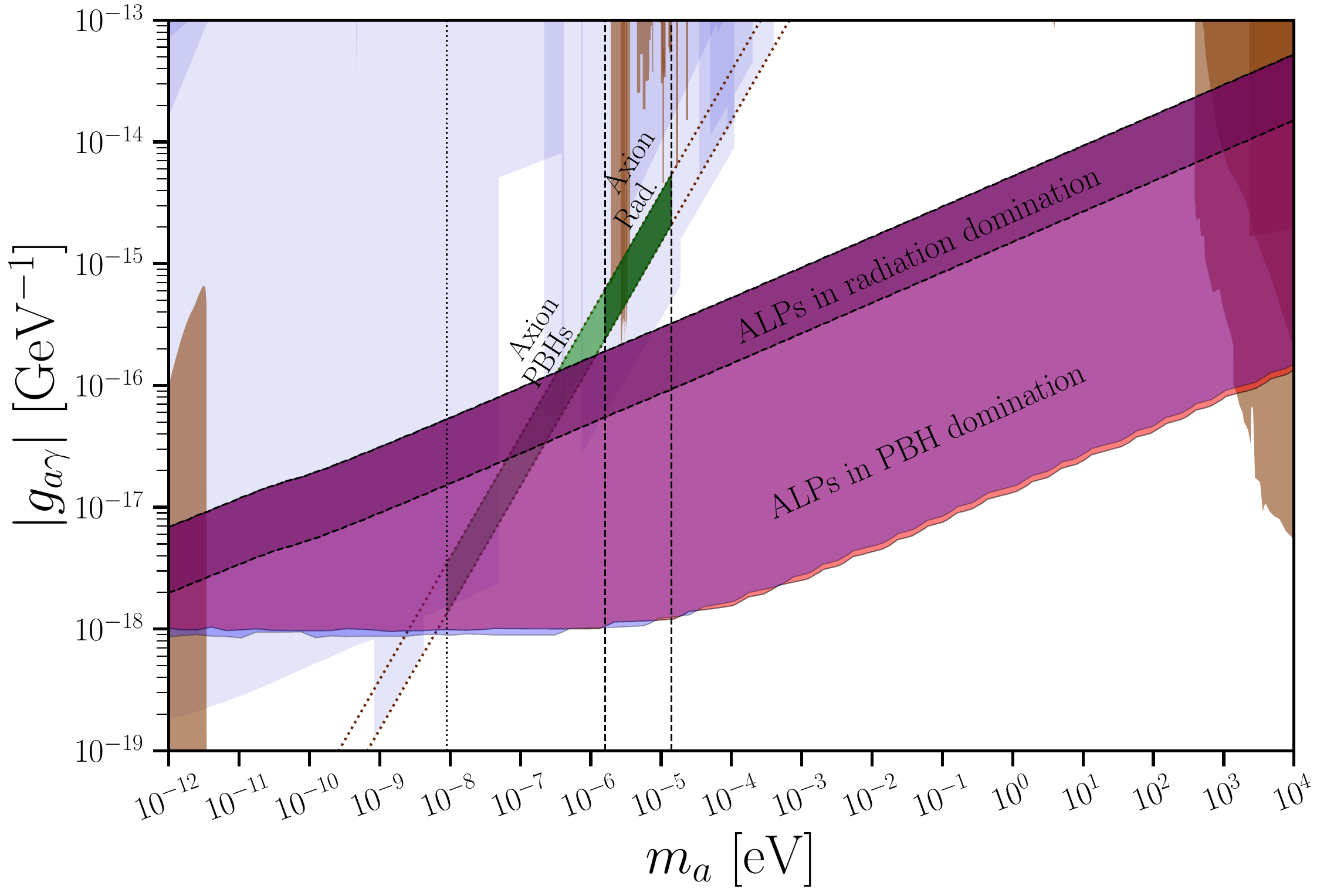}
		\caption{Viable parameter space for ALPs and the QCD axion, for radiation dominated or PBH dominated scenarios. The brown and light blue shaded regions are the current exclusions and the projected sensitivities of the different experiments described in the text.
			The figure has been adapted from Ref.~\cite{ciaran_o_hare_2020_3932430}.}
		\label{fig:bandALP}
	\end{figure}

	Once one assumes that all DM is composed by ALPs, $m_a$ and $f_a$ (and therefore $g_{a\gamma}$) are not longer independent, for a given initial misalignment angle. 
	In Fig.~\ref{fig:bandALP} we display the parameter space compatible with the DM relic density constraint, in the plane $(m_a,\, |g_{a\gamma}|)$.
	The diagonal band dubbed  ``ALPs in radiation domination'' corresponds to the viable region within the standard cosmology, with the thickness of the band representing the possible values for $\theta_i$ in the range $[0.5,\, \pi/\sqrt{3}]$, with $\theta_i = \pi/\sqrt{3}$ being the top border.
	The regions on the top and bottom give rise to a DM underabundance and overabundance, respectively.%
	\footnote{Notice that we are not considering further production mechanisms  of ALPs such as the decay of cosmic strings or domain walls~\cite{Davis:1985pt, Harari:1987ht, Battye:1994au, Hiramatsu:2012gg, Gelmini:2021yzu}.}
	This parameter space is modified once an early PBH dominated epoch occurs.
	The red (blue) thick band shows the viable parameter space for $a_\star = 0$ ($a_\star = 0.999$), enhanced by the presence of PBHs.
	The impact of the PBH spin is very mild, and therefore the two bands are almost superimposed.
	The PBH entropy injection dilutes the ALP abundance, and therefore higher values for the PQ breaking scale are allowed, which corresponds to lower values for $|g_{a\gamma}|$.
	It is interesting to note that for $m_a \lesssim 10^{-5}$~eV, $\Tosc < \Teq$, and therefore the ALP relic density is independent on its mass.
	However, for $m_a \gtrsim 10^{-5}$~eV, $\Tosc > \Teq$ and $g_{a\gamma} \propto m_a^{1/4}$, cf. Eq.~\eqref{eq:Omegah2ALP}.
	
	Figure~\ref{fig:bandALP} also overlays in brown current bounds on the ALP-photon coupling.
	Small masses ($m_a \lesssim 4 \times 10^{-11}$~eV) are constrained astrophysical black hole spins~\cite{Mehta:2020kwu}, whereas high masses ($m_a \gtrsim 10^2$~eV) by the $X$-ray background and extragalactic background light~\cite{Cadamuro:2011fd}.
	The masses in the range $10^{-6}$~eV $\lesssim m_a \lesssim 10^{-5}$~eV are already probed by the so-called haloscope experiments~\cite{Sikivie:1983ip}, ADMX~\cite{ADMX:2009iij, ADMX:2019uok}, HAYSTAC~\cite{HAYSTAC:2018rwy}, CAPP~\cite{Lee:2020cfj, Jeong:2020cwz, CAPP:2020utb}, QUAX~\cite{Alesini:2019ajt, Alesini:2020vny}, ORGAN~\cite{McAllister:2017lkb}, using a highly tuned microwave cavity that converts ALPs into photons in the presence of a static magnetic field.%
	\footnote{We note that the parameter space shown in Fig.~\ref{fig:bandALP} is consistent with an ALP lifetime larger than the age of the Universe.}
	Additionally, the light blue regions show projected sensitivities from a number of experiments that count ORGAN~\cite{McAllister:2017lkb}, MADMAX~\cite{Beurthey:2020yuq}, ALPHA~\cite{Lawson:2019brd}, ADMX~\cite{Stern:2016bbw}, KLASH~\cite{Alesini:2017ifp}, DM-Radio~\cite{Chaudhuri:2018rqn} and ABRACADABRA~\cite{Kahn:2016aff, Ouellet:2018beu}.
	These bounds have been adapted from Ref.~\cite{ciaran_o_hare_2020_3932430}.
	Even if the PBH domination tends to imply smaller values for $|g_{a\gamma}|$, this parameter space could be potentially tested in the future by next-generation experiments, especially for light ALP masses, $m_a \lesssim 10^{-7}$~eV. 
	
	Before closing this section, it is interesting to note that further constraints on $f_a$ appear depending on the scale at which the spontaneous breakdown of the PQ symmetry happened.
	If the  PQ symmetry was broken during inflation ($f_a > H_I$, with $H_I$ being the inflationary scale) and never restored, the ALP field would be homogenized through the Hubble patch.
	On the other hand, if the inflationary epoch ends before the PQ transition  ($f_a < H_I$) the initial misalignment angle would vary along the different patches of the Universe. 
	Let us consider in more detail these two scenarios and their possible relation with PBHs.
	
	\subsubsection*{Pre-inflationary scenario}
	In this case, the PQ symmetry is spontaneously broken during inflation (i.e. $H_I < f_a$), and it is not restored afterwards~\cite{Dine:1982ah}. 
	The ALP field is homogeneous through various Hubble patches, with a unique value of $\theta_i$ characterizing the whole observable Universe.
	In this scenario, the ALP is present during inflation and therefore ALP isocurvature fluctuations (converted into curvature perturbations) are expected to leave a imprint on the CMB. However, since the CMB measurements do not allow for sizable isocurvature modes, the scale of inflation is pushed to relatively low values.
	Accordingly, the isocurvature bounds obtained from Planck data~\cite{Planck:2018jri} impose the lower bound on $f_a$ which can be cast as~\cite{DiLuzio:2020wdo}
	\begin{equation} \label{eq:Hiso-alp}
	H_I \lesssim 0.9 \times 10^7~\mathrm{GeV} \left(\frac{\theta_i}{\pi}\right) \left(\frac{f_a}{10^{11}~\mathrm{GeV}}\right).
	\end{equation}
	Since the BBN scale represents a lower bound for $H_I$, it follows that $f_a\gtrsim100$ GeV  ($|g_{a\gamma}| \lesssim 10^{-5}$~GeV$^{-1}$) for $\theta_i \sim \mathcal{O}(1)$, which in turn  does not have an impact on the parameter space displayed in Fig.~\ref{fig:bandALP}.   
	A stronger limit appears when examining the connection with PBHs.
	Combining the above expression with Eq.~\eqref{eq:Mi} we obtain a lower limit for the initial mass of the PBH
	\begin{equation}
	\Min = 4\pi\, \gamma\, \frac{M_P^2}{H(\Tin)}
	\gtrsim 4\pi\, \gamma\, \frac{M_P^2}{H_I}
	\simeq 8.6 \times 10^6~\mathrm{g} \left(\frac{1}{\theta_i}\right) \left(\frac{10^{11}~\mathrm{GeV}}{f_a}\right),
	\end{equation}
	which reflects the fact that PBHs are created after inflation, in a radiation-dominated epoch.
	For PBHs fully evaporating before the onset of BBN, $\Min \lesssim 6 \times 10^{8}$~g, and therefore for $\theta_i \sim \mathcal{O}(1)$, it implies that
	\begin{equation}
	|g_{a\gamma}| \lesssim 10^{-12}~\textrm{GeV}^{-1},
	\end{equation}
	which is again automatically satisfied for our relevant parameter space.

	\subsubsection*{Post-inflationary scenario}
	For this case, the PQ symmetry is spontaneously broken after the inflationary period, i.e. $f_a < H_I$. 
	The misalignment angle takes random values along different Hubble patches, in which case it is averaged out over many patches (in the range $[-\pi,\, \pi)$) so that $\theta_i = \sqrt{\langle\theta_i^2\rangle} =\pi/\sqrt{3}$.%
	\footnote{When taking into account the anharmonicity of the potential, the average becomes $\theta_i = \sqrt{\langle\theta_i^2\rangle} \simeq 2.15$~\cite{GrillidiCortona:2015jxo}.}
	Additionally, the upper limit on the inflationary scale $H_I < 2.5 \times
	10^{-5}~M_P$~\cite{Planck:2018jri} implies that $f_a \lesssim 10^{13}$~GeV (and in turn $g_{a\gamma} \gtrsim 10^{-16}$~GeV$^{-1}$), which finally can be translated into an lower bound on the ALP mass $m_a \gtrsim 10^{-7}$~eV, for ALPs being the whole DM.
	
	An important feature of the post-inflationary scenario is that the value of the initial misalignment angle changes by a factor of $\mathcal{O}(1)$ from one causal patch to the next.
	Accordingly, the density of cold ALPs is characterized by sizable inhomogeneities.
	Their free streaming length is too short to erase these inhomogeneities before the matter-radiation equality, so that the density perturbations decouple from the Hubble expansion and start growing by gravitational instability, rapidly forming gravitationally bound objects, called miniclusters~\cite{Hogan:1988mp, Kolb:1993zz, Kolb:1993hw, Kolb:1994fi,  DiLuzio:2020wdo}. 
	
\begin{figure}
		\def\sepf{0.57}
		\centering
		\includegraphics[scale=\sepf]{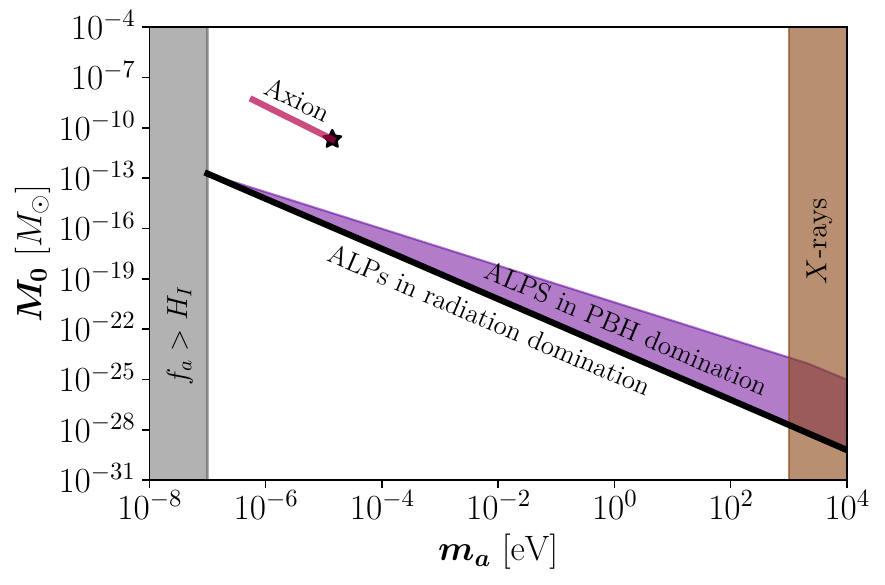}
		\caption{Minicluster mass in the post-inflationary scenario, at the time of formation.
			ALPs: The thick black line corresponds to the the standard cosmological scenario, whereas the purple band to the PBH domination.
			Axions: The star corresponds to the the standard cosmological scenario, whereas the purple line to the PBH domination.
			The vertical bands are excluded because $f_a > H_I$, or in tension with $X$-rays measurements.
		}
		\label{fig:M0-ALP}
	\end{figure}

	The formation time of ALP miniclusters is sensitive to the cosmology prior to BBN and, in particular, to the temperature at which ALP oscillations take place.
	The scale of ALP minicluster mass is set by the total mass of ALPs within one Hubble volume of radius $R_\text{osc} \sim H(\Tosc)^{-1}$, at the time of the matter-radiation equality. Therefore, their mass at formation is
	\begin{equation} \label{eq:mc}
	M_0 = \frac{4\pi}{3}\, \rho_{\rm DM}(T_0)\, \frac{s(\Tosc)}{s(T_0)} \left(\frac{1}{H(\Tosc)}\right)^3 \times \frac{S(\Tev)}{S(\Tosc)}\,.
	\end{equation}
	In a radiation dominated Universe and assuming that all relic ALPs bound up in miniclusters, it reduces to
	\begin{equation}
	M_0 = \frac{8\, \gss(\Tosc)}{5} \left(\frac{10}{\gs(\Tosc)}\right)^{3/2} \frac{\Omega_{\rm DM} h^2\, \rho_c/h^2}{s(T_0)}\, \frac{M_P^3}{\Tosc^3}
	\simeq 2 \times 10^{-16} \left(\frac{10^{-5}~\text{eV}}{m_a}\right)^{3/2} M_\odot\,,
	\end{equation}
	as shown in Fig.~\ref{fig:M0-ALP} with a solid black line.
	The vertical bands are excluded because $f_a > H_I$, or in tension with the $X$-ray background and the extragalactic background light~\cite{Cadamuro:2011fd}.
	
	However, if ALPs start to oscillate during the PBH dominated period, the initial minicluster mass can be estimated to be%
	\footnote{It is worth mentioning that in either a radiation or early matter domination era may be possible the formation of ALP miniclusters even in pre-inflationary scenarios~\cite{Hardy:2016mns,Nelson:2018via}.}
	\begin{equation}
	M_0 \simeq \frac{16 \pi\, \gss}{\gs}\, \frac{\Omega_{\rm DM} h^2\, \rho_c/h^2}{s(T_0)} \times
	\begin{dcases}
	\pi^{1/2} \left(\frac{\gs}{10}\right)^{1/4} \left(\frac{M_P}{m_a}\right)^{3/2} \frac{\Teq}{\Tev} & \text{for } \Tosc \geq \Teq\,,\\
	\frac{M_P^2}{m_a\, \Tev} & \text{for } \Teq \geq \Tosc \geq \Tev\,,\\
	\pi^{1/2} \left(\frac{\gs}{10}\right)^{1/4} \left(\frac{M_P}{m_a}\right)^{3/2} & \text{for } \Tev \geq \Tosc\,.
	\end{dcases}
	\end{equation}
	Figure~\ref{fig:M0-ALP} also shows the maximal impact of the PBH domination era with a broad purple band.
	Again, we have taken the PBH benchmark values of $\Min \simeq 5.7 \times 10^8$~g and $\beta \simeq 4.6 \times 10^{-10}$ for Schwarzschild, or $\Min \simeq 7.6 \times 10^8$~g and $\beta \simeq 3.5 \times 10^{-10}$ for Kerr.
	The spin of the PBHs has a very limited impact, and therefore the differences in the plot are barely visible.
	However, it is clear that heavier ALP miniclusters can be formed due to the entropy injection. In other words, the entropy injection works as an {\it enhancement} factor, cf. Eq.~\eqref{eq:mc} (see e.g. Refs.~\cite{Nelson:2018via, Visinelli:2018wza, Blinov:2019jqc} for related studies).
	Hence, searches for ALPs, such as indirect detection through gravitational microlensing~\cite{Kolb:1995bu, Zurek:2006sy, Fairbairn:2017dmf, Fairbairn:2017sil}, could be extended to regions of the parameter space previously thought to be not relevant.
	
	\section{Axion Dark Matter} \label{axion_DM}
	The production of QCD axion in a cosmology dominated by PBHs was analytically studied in Ref.~\cite{Bernal:2021yyb}.
	For the sake of completeness, here we revisit that scenario from a numerically perspective. In particular, we: $i)$ include the greybody factors in the PBHs emission properties, $ii)$ take the effect of the PBH angular momentum into account, and $iii)$ numerically solve the evolution of the PBH and radiation energy density.
	This is particularly pertinent because it allows to take into account the time-dependent BH evaporation rate.
	
\begin{figure}
		\def\sepf{0.57}
		\centering
		\includegraphics[scale=\sepf]{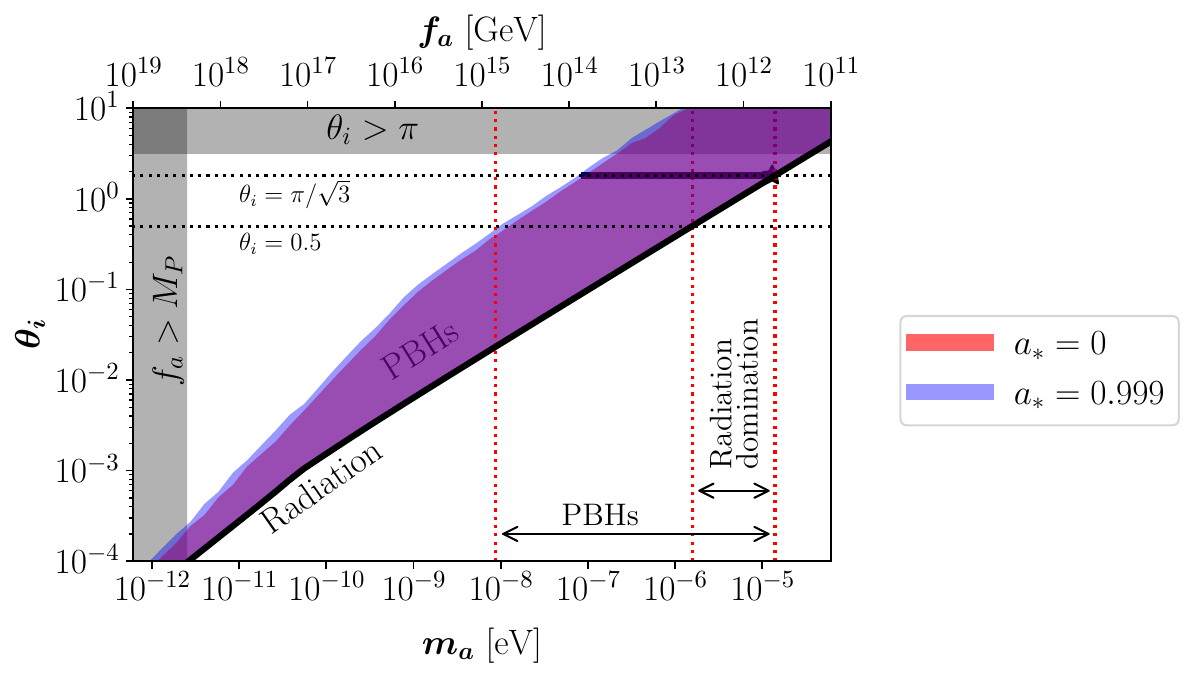}
		\caption{Initial misalignment angle of QCD axions required to explain the observed DM abundance for $a_\star=0$ (red band) and $a_\star=0.999$ (blue band). The corresponding values in the standard cosmological scenario are represented by the black thick line. 
		}
		\label{fig:band}
	\end{figure}

	For the case of QCD axions, the scale $f_a$ at which the PQ $U(1)$ symmetry is spontaneously broken determines the axion mass through the topological susceptibility of QCD~\cite{Borsanyi:2016ksw}
	\begin{equation}
	\chi(T) \simeq 0.0245~\text{fm}^{-4} \times 
	\begin{dcases}
	1 &\text{ for } T \leq \Tqcd\,,\\
	\left(\frac{T}{\Tqcd}\right)^{-8.16} &\text{ for } T \geq \Tqcd\,,
	\end{dcases}
	\end{equation}
	as
	\begin{equation}
	m_a(T) = \frac{\sqrt{\chi(T)}}{f_a}\,,
	\end{equation}
	with $\Tqcd \simeq 150$~MeV.
	As the axion mass is temperature dependent, its abundance in the standard cosmology is
	\begin{equation} \label{eq:QCDaxion}
	\Omega_ah^2 \simeq 0.12 \left(\frac{\theta_i}{10^{-3}}\right)^2
	\times
	\begin{dcases}
	\left(\frac{m_a}{m_a^\text{QCD}}\right)^{-\frac32} & \text{for } m_a \leq m_a^\text{QCD},\\
	\left(\frac{m_a}{m_a^\text{QCD}}\right)^{-\frac76} & \text{for } m_a \geq m_a^\text{QCD},
	\end{dcases}
	\end{equation}
	with $m_a^\text{QCD} \equiv m_a(\Tosc=\Tqcd) \simeq 4.8 \times 10^{-11}$~eV.
	However, if a PBH-dominated epoch occurs, the axion relic abundance gets modified due to the reduction of its oscillation temperature and its dilution induced by the entropy injection~\cite{Venegas:2021wwm, Bernal:2021yyb, Arias:2021rer}. Again, we numerically solve the coupled Boltzmann equations, determine the dilution factor and further calculate the modified QCD axion relic density. In Fig.~\ref{fig:band}, we present our numerical results, which  agrees well with the analytical estimations presented in Ref.~\cite{Bernal:2021yyb}.
	The purple and blue bands of Fig.~\ref{fig:band}  show the misalignment angle required to reproduce the whole observed axion DM abundance for the case with Schwarzschild ($a_\star=0$) and  Kerr ($a_\star=0.999$) PBH domination, respectively. The black thick line corresponds to the standard cosmological scenario. The thickness of the band brackets all possible PBH scenarios compatible with BBN ($\Tev > 4$~MeV) and GWs (cf. Eq.~\eqref{eq:GW}).
	Once we restrict to initial misalignment angles in the range $\theta_i \in [0.5,\, \pi/\sqrt{3}]$, we have that the axion mass can take lower values than those allowed in a purely radiation dominated epoch. Concretely, the axion mass range gets broaden towards low values from $\sim [10^{-6},\, 10^{-5}]$~eV to $\sim [10^{-8},\, 10^{-5}]$~eV.     
	This enlargement of the viable mass range is also shown in Fig.~\ref{fig:bandALP}, now taking into account the axion coupling to photons.
	The thickness of the band correspond to the KSVZ and DFSZ models for the axion-photon coupling.
	It is interesting to see that this parameter space is within the reach of future detectors as ABRACADABRA, KLASH, ADMX, and DM-Radio.
	
	\subsubsection*{Pre-inflationary scenario}
	When axion is present during inflation (and the PQ symmetry is not restored after it), the requirement of axions to constitute all DM along with the standard cosmological scenario along with the isocurvature bound~\eqref{eq:Hiso-alp} leads to
	\begin{equation} \label{eq:Hiso-axion}
	H_I \lesssim
	\begin{dcases}
	3.2 \times 10^{9}~\mathrm{GeV} \left(\frac{f_a}{10^{17}~\mathrm{GeV}}\right)^{1/4} & \text{ for } f_a\gtrsim 1.2 \times 10^{17}~\text{GeV},\\
	2.6 \times 10^{7}~\mathrm{GeV} \left(\frac{f_a}{10^{12}~\mathrm{GeV}}\right)^{5/12}  & \text{ for } f_a\lesssim 1.2 \times 10^{17}~\text{GeV}.
	\end{dcases}
	\end{equation}
	Consequently, under the PBH domination slightly higher inflation scales can be reached since $f_a$ is allowed to take higher values. In other words, the relevant parameter space displayed in Figs.~\ref{fig:bandALP} and~\ref{fig:band} does not get affected.                         
	Translating the above bounds into the initial mass for the PBHs, we have    
	\begin{equation}
	\Min \gtrsim
	\begin{dcases} 
	7.5 \times10^{3}~\mathrm{g} \left(\frac{10^{17}~\mathrm{GeV}}{f_a}\right)^{1/4} & \text{ for } f_a\gtrsim 1.2 \times 10^{17}~\text{GeV},\\
	9.4 \times10^{5}~\mathrm{g} \left(\frac{10^{12}~\mathrm{GeV}}{f_a}\right)^{5/12}  & \text{ for } f_a\lesssim 1.2 \times 10^{17}~\text{GeV}.
	\end{dcases}
	\end{equation}
	Contrary to the ALP case, there is no significant lower bound on $f_a$ when $\Min$ approaches to the maximum value allowed by BBN. 
	
	\subsubsection*{Post-inflationary scenario}
	Since the PQ symmetry remains conserved during the inflation period ($f_a < H_I$), the axion field does not get homogenized.  The upper limit on the inflationary scale implies that $f_a \lesssim 10^{13}$~GeV, and hence the lower bound on the axion mass $m_a \gtrsim 6 \times 10^{-7}$~eV.
	
	In the standard cosmological scenario, axions start oscillating at the temperature $\Tosc \simeq 1.1$~GeV. Therefore, if the misalignment mechanism is the unique method to produce axions, they should have a mass $m_a \simeq 1.8 \times 10^{-5}$~eV (for $\theta_i=\pi/\sqrt{3}$) in order to account for the whole of the DM abundance. This case is identified by a star in the upper right region in the Fig.~\ref{fig:band}. 
	This result is remarkably modified by the dilution factor $S(\Tosc)/S(\Tev)$ that emerges within the PBH domination, which now allows for a mass range instead of a single point, reaching values up to two orders of magnitude lower, i.e. $\mathcal{O}(10^{-7})$~eV. This new range is represented by the thick solid horizon line in Fig.~\ref{fig:band}. 
	
	The gravitational clump of axion density inhomogeneities at the time of radiation-matter equality leads to 
	miniclusters having a mass $M_0 \simeq 2.1 \times 10^{-11}~M_\odot$ in the standard cosmological scenario (see the star mark in Fig.~\ref{fig:M0-ALP}). 
	However, in a PBH dominated scenario, such a mass gets enhanced due to the entropy injection by BH evaporation.
	The mass $M_0$ of the axion miniclusters at formation becomes
	\begin{equation}
	M_0 \simeq \frac{16\, \gss}{\gs}\, \frac{\Omega h^2\, \rho_c/h^2}{s(T_0)} \times
	\begin{dcases}
	\pi^{1/2} \left(\frac{10}{\gs}\right)^{1/4} \frac{M_P^{5/2}}{\Tqcd^2\, m_a^{1/2}} \frac{\Teq}{\Tev} & \text{for } \Tosc \geq \Teq\,,\\
	\pi^{3/11} \left(\frac{10}{\gs}\right)^{4/11} \frac{M_P^{30/11}}{\Tqcd^{12/11}\, m_a^{3/11}\, \Teq^{4/11}\, \Tev} & \text{for } \Teq \geq \Tosc \geq \Tc\,,\\
	\pi^{1/2} \left(\frac{10}{\gs}\right)^{1/4} \frac{M_P^{5/2}}{\Tqcd^2\, m_a^{1/2}} & \text{for }  \Tc \geq \Tosc\,.
	\end{dcases}
	\end{equation}
	In Fig.~\ref{fig:M0-ALP} we show  the enhancement on $M_0$ (purple line) generated by the PBH domination. It is remarkable that now $M_0$ can reach values as high as $10^{-8}~M_{\odot}$.

	\section{Dark Radiation} \label{axion_DR}
	Axions and ALPs emitted from PBH evaporation are ultra-relativistic, thus behaving as DR and contributing to the effective number of neutrinos~\cite{Hooper:2019gtx, Lunardini:2019zob, DiLuzio:2020wdo, Schiavone:2021imu}.
	We note that ALP DR directly emitted from nonrotating PBHs was analyzed in Ref.~\cite{Schiavone:2021imu} recently. Here we revisit  ALP DR from PBHs with a more general perspective, namely we will study both nonrotating and rotating PBHs.
	The contribution to effective number of neutrinos can be simply estimated, obtaining
	\begin{equation}
	\Delta N_{\text{eff}} = \left[\frac87 \left(\frac{11}{4}\right)^{\frac43} + N_{\text{eff}}^{\text{SM}}\right] \frac{\rho_a(\Tev)}{\rR(\Tev)} \left(\frac{\gs(\Tev)}{\gs(\Tequal)}\right) \left(\frac{\gss(\Tequal)}{\gss(\Tev)}\right)^\frac43,
	\end{equation}
	where $\Tequal \simeq 0.75$~eV denotes the temperature at (late) matter-radiation equality. 
	In the PBH-dominated scenario, the ratio between the axion (or ALP) and the radiation energy densities is related to their contributions to the evaporation function
	\begin{equation}
	\frac{\rho_a(\Tev)}{\rR(\Tev)} = \frac{\varepsilon_a(z_i,a_\star)}{\varepsilon_R(z_i,a_\star)}\,,
	\end{equation}
	with $\varepsilon_a(\Tev)$ being the axion evaporation function, and $\varepsilon_R(\Tev)$ the total SM radiation contribution to the evaporation function. For the case of Schwarzschild PBHs, $\Delta N_{\rm eff}$ is simply
	\begin{equation}\label{eq:DNeffd}
	\Delta N_\text{eff} \simeq 0.237\, \frac{\gs\left(\Tev\right)}{\gss\left(\Tev\right)^\frac43}\,,
	\end{equation}
	which is independent from $\beta$, and consistent with the result obtained in Ref.~\cite{Schiavone:2021imu}.
	For Kerr PBHs, the time depletion of the angular momentum makes it more difficult to obtain a simple analytical form. 
	
	\begin{figure}
		\def\sepf{0.51}
		\centering
		\includegraphics[scale=\sepf]{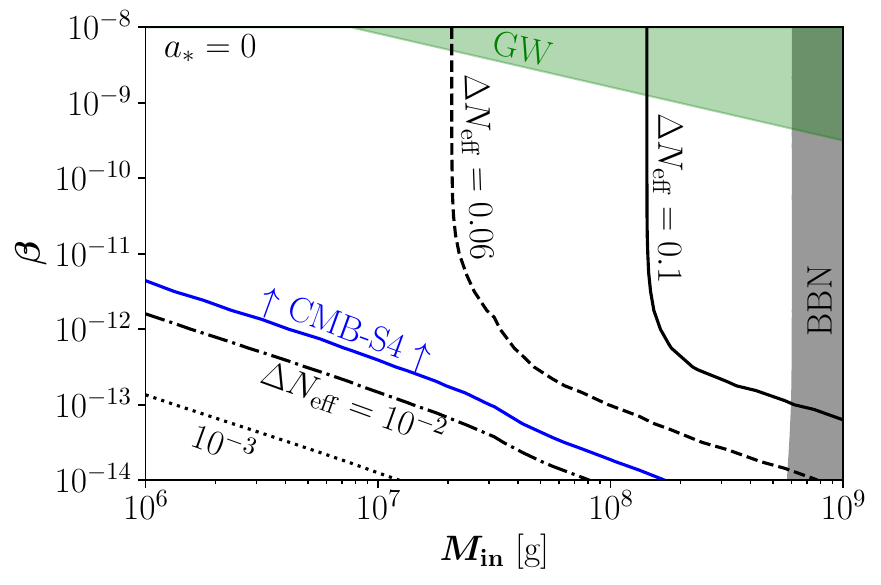}
		\includegraphics[scale=\sepf]{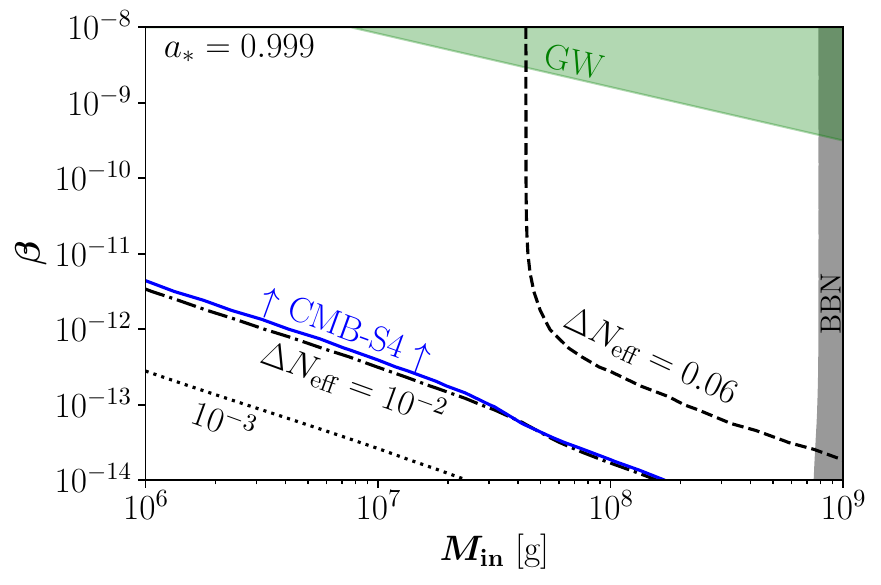}
		\caption{Contribution to the effective number of neutrinos $\Delta N_\text{eff}$ coming from axions/ALPs directly emitted via Hawking radiation for $a_\star = 0$ (left) and $a_\star = 0.999$ (right). The blue line indicates the region within the reach of the future CMB-S4 experiment~\cite{CMB-S4:2016ple}.
		}
		\label{fig:DNeff}
	\end{figure} 
	Left (right) panel of Fig.~\ref{fig:DNeff} shows the contribution to $\Delta N_\text{eff}$ as function of BH mass and $\beta$ for Schwarzschild (Kerr) PBHs. 
	Let us note that in a PBH-dominated era, $\Delta N_{\rm eff}$ tends to be independent of $\beta$, as noted before, c.f. Eq.~\eqref{eq:DNeffd}.
	On the contrary, when $\beta < \beta_c$,  larger $\Min$ is needed with smaller $\beta$ in order to yield a fixed $\Delta N_{\text{eff}}$.
	We also note that when $\Min \simeq 2 \times 10^{7}$~g (correspondingly $\Tev \simeq 0.1~\text{GeV}$), there are some features in the lines shown in Fig.~\ref{fig:DNeff}. This is because the sharp change of $\gs$ and $\gss$ due to QCD phase transition~\cite{Hooper:2019gtx, Lunardini:2019zob}.
	Comparing the resulting values of $\Delta N_{\rm eff}$ for Schwarzschild and Kerr PBH, we observe that for the latter case the final values are smaller than those for nonrotating PBHs.
	This is simply related to the properties of Kerr PBHs: if the BH spin is close-to-maximal, the emission of higher spin particles is enhanced in the early stages of the evaporation~\cite{Page:1976df}.
	Thus, the particle injection to SM radiation density is larger than in the Schwarzschild case.
	Meanwhile, the axion/ALP emission is not affected by the BH spin, so their final contribution to $\Delta N_{\rm eff}$ is in fact reduced~\cite{Hooper:2020evu, Masina:2021zpu}.
	The current upper bound on $\Delta N_\text{eff}$ comes from the Planck collaboration, and reads $N_{\mathrm{eff}} = 2.99 \pm 0.17$~\cite{Planck:2018vyg}.  
	However, $\Delta N_\text{eff}$ due to emitted hot axions/ALPs could reach the sensitivity of next generation CMB experiments: CMB-S4~\cite{CMB-S4:2016ple} ($\Delta N_{\mathrm{eff}}\sim 0.06$), PICO~\cite{NASAPICO:2019thw} ($\Delta N_{\mathrm{eff}}\sim 0.06$) and SPT-3G/SO~\cite{Benson:2014qhw,Ade:2018sbj} ($\Delta N_{\mathrm{eff}} \sim 0.1$), and therefore axion/ALP DR could be tested in the near future.%
	\footnote{It is interesting to note that extra radiation, at the level of  $0.2 \lesssim \Delta N_\text{eff} \lesssim 0.5$, can alleviate the tension between measurements of the Hubble parameter at early and late times~\cite{Riess:2016jrr, Planck:2018vyg, Hooper:2019gtx, Escudero:2019gzq, Vagnozzi:2019ezj, Alcaniz:2019kah}.
		However, such large levels can not be reached by PBHs.}
	
	\section{Conclusions} \label{conclusion}
	
	PBHs are without doubt one of the most interesting objects that could exist in nature.
	As the PBH energy density scale like non-relativistic matter, they can naturally dominate the expansion rate of the Universe, triggering a nonstandard cosmological epoch.
	Moreover, they also source all types of particles, thus injecting large quantities of entropy to the primordial plasma.
	Hence, these two specific properties, namely, the possibility of dominating expansion of the Universe and the large injection of entropy, make the phenomenology of a PBH dominated Universe unique among the possible scenarios for a nonstandard cosmology.  
	
	In this paper, we studied the phenomenological consequences of axion-like particles (ALPs) as dark matter candidates in an early PBH-dominated epoch.
	To that end, we numerically solved the set of Boltzmann equations for the background dynamics, carefully taking into account both the greybody factors and the PBH angular momentum.
	PBHs have a strong impact on the misalignment ALP production.
	On the one hand, as the Hubble expansion rate is enhanced, the oscillation temperature decreases, which corresponds to a delay in the oscillations.
	On the other hand, the entropy injection due to the PBH evaporation has to be compensated by a larger spontaneous breaking scale value of the Peccei-Quinn symmetry or, equivalently, by a smaller ALP-photon coupling.
	An equivalent trend happens for QCD axions, with the particularity that its standard mass window $m_a \simeq [10^{-6}-10^{-5}]$~eV gets broaden to $\sim [10^{-8}-10^{-5}]$~eV (for misalignment angles of order $\mathcal{O}(1)$), mainly due to the entropy injection.
	It is interesting to note that both for ALPs and axions, the new viable parameter space is within the projected sensitivities of detectors as ABRACADABRA, KLASH, ADMX, and DM-Radio.
	
	Additionally, we showed that at production the ALP/axion minicluster mass could increase by several orders of magnitude, due to the enhancement required by the PBH entropy injection.
	This could give new prospects for the indirect detection via gravitational microlensing of ALP and axion dark matter, with respect to the standard cosmological scenario.
	
	Finally, we analyzed in detail the dark radiation arising from relativistic axions and ALPs directly emitted from PBHs evaporation.
	For Kerr PBH, the contribution $\Delta N_{\text{eff}}$ are smaller than those for Schwarzschild PBHs.
	Interestingly, this contribution is within the projected reach of future CMB Stage 4 experiments and could help relaxing the tension between late and early-time Hubble determinations.

	\section*{Acknowledgments}
	NB received funding from the Spanish FEDER/MCIU-AEI under grant FPA2017-84543-P, and the Patrimonio Autónomo - Fondo Nacional de Financiamiento para la Ciencia, la Tecnología y la Innovación Francisco José de Caldas (MinCiencias - Colombia) grant 80740-465-2020.
	The work of OZ is supported by Sostenibilidad-UdeA and the UdeA/CODI Grants 2017-16286 and 2020-3317.
	This project has received funding/support from the European Union's Horizon 2020 research and innovation programme under the Marie Skłodowska-Curie grant agreement No 860881-HIDDeN. 
	Fermilab is operated by the Fermi Research Alliance, LLC under contract No. DE-AC02-07CH11359 with the United States Department of Energy.
	
	\bibliographystyle{JHEP}
	\bibliography{biblio}
	
\end{document}